\documentclass[twocolumn,preprintnumbers,nofootinbib,prc,superscriptaddress]{revtex4-1}

\usepackage{graphicx}
\usepackage{amssymb}
\usepackage{epstopdf}
\usepackage{dcolumn}
\usepackage{bm}

\begin{document}
    
\title{Comments on Systematic Effects in the NIST Beam Neutron Lifetime Experiment}
    
\author{F.~E.~Wietfeldt}
\affiliation{Department of Physics and Engineering Physics, Tulane University, New Orleans, LA 70118, USA}
\author{R.~Biswas}
\affiliation{Department of Physics and Engineering Physics, Tulane University, New Orleans, LA 70118, USA}
\author{J.~Caylor}
\affiliation{Department of Physics, University of Tennessee, Knoxville, TN 37996, USA}
\author{B.~Crawford}
\affiliation{Physics Department, Gettysburg College, Gettysburg, PA 17325, USA}
\author{M.~S.~Dewey}
\affiliation{National Institute of Standards and Technology, Gaithersburg, MD 20899, USA}
\author{N.~Fomin}
\affiliation{Department of Physics, University of Tennessee, Knoxville, TN 37996, USA}
\author{G.~L.~Greene}
\affiliation{Department of Physics and Engineering Physics, Tulane University, New Orleans, LA 70118, USA}
\affiliation{Department of Physics, University of Tennessee, Knoxville, TN 37996, USA}
\author{C.~C.~Haddock}
\affiliation{National Institute of Standards and Technology, Gaithersburg, MD 20899, USA}
\author{S.~F.~Hoogerheide}
\affiliation{National Institute of Standards and Technology, Gaithersburg, MD 20899, USA}
\author{H.~P.~Mumm}
\affiliation{National Institute of Standards and Technology, Gaithersburg, MD 20899, USA}
\author{J.~S.~Nico}
\affiliation{National Institute of Standards and Technology, Gaithersburg, MD 20899, USA}
\author{W.~M.~Snow}
\affiliation{Physics Department, Indiana University, Bloomington, IN 47405, USA}
\author{J.~Zuchegno}
\affiliation{Department of Physics and Engineering Physics, Tulane University, New Orleans, LA 70118, USA}

\date{\today}

\begin{abstract}
We discuss issues raised by Serebrov, {\em et al.} in a recent paper \cite{Ser21} regarding systematic effects in the beam neutron lifetime experiment performed at NIST \cite{Dew03,Nic05,Yue13}. We show that these effects were considered in the original analyses and that our corrections and systematic uncertainties were appropriate. We point out some misconceptions and erroneous assumptions in the analysis of Serebrov, {\em et al}. None of the issues raised in Ref. \cite{Ser21} lead us to alter the value of the neutron lifetime reported in Ref. \cite{Yue13}.
\end{abstract}

\maketitle

\section{Introduction}
\label{S:Intro}

The earliest measurements of the free neutron decay lifetime, beginning with Robson's landmark 1951 experiment \cite{Rob51}, were made by counting decay products (protons and/or electrons) from a thermal neutron beam, while simultaneously measuring the neutron beam density  using a thin, neutron-absorbing foil. The neutron lifetime $\tau_n$ is found from the differential form of the exponential decay law: $ \tau_n = N/\dot{N}$, where $N$ is the number of neutrons in the decay volume and $\dot{N}$ is the observed decay rate, corrected for detection efficiencies. This approach is often called the {\em beam method}. Starting in the 1980s ultracold neutrons (UCN) with kinetic energy $\approx 10^{-7}$ eV were produced and stored for long periods in material bottles, enabling a new method to measure the neutron lifetime: repeatedly filling the bottle and counting the UCN that remain after variable storage times. This is generally called the {\em UCN storage method} or simply the {\em bottle method}. A detailed review of the methods and history of neutron lifetime measurements can be found in Ref. \cite{Wie11}. A  recent innovation in the UCN storage method is to confine polarized UCN in a magnetic bottle \cite{Osh09, Ezh18, Pat18, Gon21} thereby eliminating the effects of neutron loss by absorption and scattering from a material surface. The most precise reported neutron lifetime measurement to date, UCN$\tau$ 2021 \cite{Gon21}, used this approach. In the past few years a third method for measuring the neutron lifetime, using spacecraft-based neutron detectors to count the relative neutron flux as a function of altitude above  Mercury, Venus, and the Moon, has reported impressive new results \cite{Wil20,Wil21} but its precision is not yet competitive.
\par
At different times over the past 60 years, experimental neutron lifetime results have been either in good or poor agreement. Currently the agreement is poor. In particular the value reported by the most precise beam experiment conducted at the NIST Center for Neutron Research \cite{Dew03,Nic05,Yue13} is 8.9 s (3.9 standard deviations) higher than the average UCN storage value using material and magnetic bottles. Other beam method results are similarly higher but with larger uncertainties. This discrepancy has been widely discussed in recent years in both the scientific literature and popular media. Due to its higher reported precision compared to other beam measurements, the NIST experiment plays a key role here. One or more unaccounted systematic effects in that result could effectively explain the discrepancy, so it has justifiably been subject to scrutiny by both the experimental team and by other scientists.
\par
In a recent paper, Serebrov, {\em et al.} \cite{Ser21} discuss and analyze three potential systematic effects in the NIST experiment: 1) protons missing the active area of the proton detector; 2) losses due to the detector dead layer; and 3) residual gas effects. We note that the authors of \cite{Ser21} based their work on what was written and published in \cite{Dew03,Nic05,Yue13} but did not seek additional details from us in advance of their publication. Here we respond to the analysis and conclusions in \cite{Ser21} and address several errors and misconceptions therein.

\section{Protons Missing the Detector}
\label{S:pMiss}
\begin{figure}
\centering
\includegraphics[width = 3.5in]{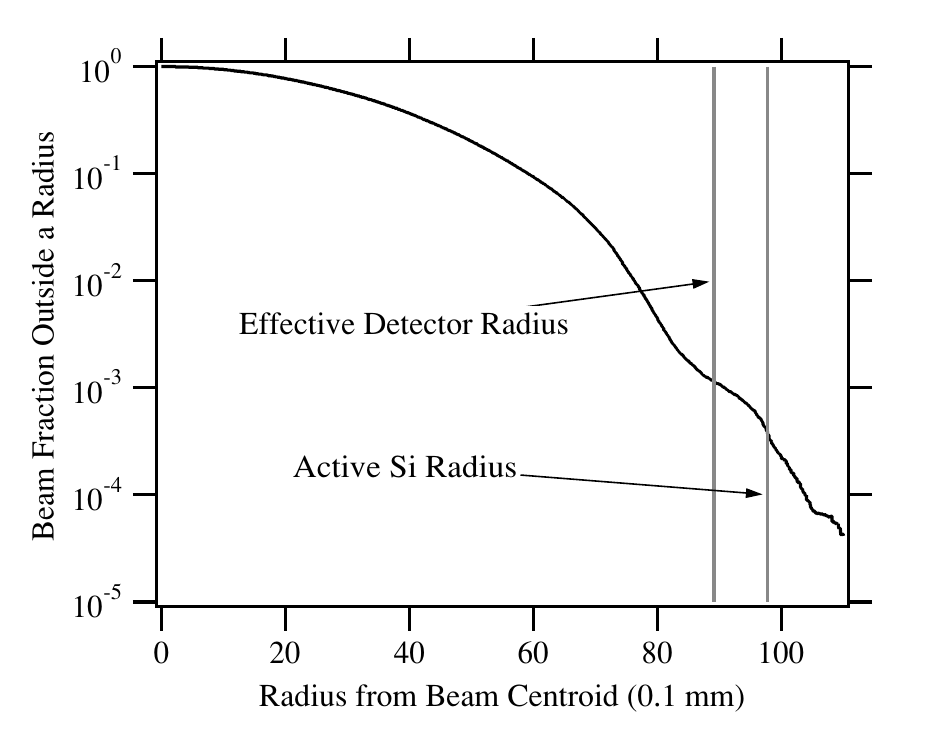}
\caption{\label{F:BL1fig11} The measured fraction of neutrons outside a given radius, based on beam images made using the dysprosium foil method. Reprinted from Ref. \cite{Nic05} (figure 11). The ``Effective Detector Radius'' accounts for the proton cyclotron orbits.}
\end{figure}

The first question Serebrov, {\em et al.} consider is whether all trapped neutron decay protons will strike the active region of the detector when the trap is opened for counting. This  was already carefully addressed in the experiment and analysis as described in Ref. \cite{Nic05}. Neutron beam intensity images were made at various positions using the dysprosium foil method. The images taken 10 cm downstream of the last trap electrode were used to estimate the proton distribution in the trap. The beam expanded gradually as it passed through the trap so these images slightly overestimated the radius of the proton distribution. Figure 11 in Ref. \cite{Nic05}, reproduced as figure \ref{F:BL1fig11} here, shows the integral of this neutron beam image, corrected for blooming effects in the dysprosium method, compared to the 9.8 mm active radius of the proton detector. The ``Effective Detector Radius'' in figure \ref{F:BL1fig11} accounts for the proton cyclotron orbit.  We concluded that $<1.1 \times 10^{-3}$ of protons will miss the detector, an upper limit due to the beam expansion, which implies a correction of -1.0 s or less. We assigned a large uncertainty, 1.0 s, to this estimate. The ``Neutron beam halo'' systematic correction of \mbox{($-1.0 \pm 1.0$) s}\footnote{All uncertainties given are 1 sigma.} listed in table V of Ref. \cite{Nic05} includes the effects of the cyclotron orbit, the blooming artifact, and the neutron beam distribution in figure \ref{F:BL1fig11}.
Serebrov, {\em et al.} essentially repeat this estimate, but without the benefit of the beam image data\footnote{Raw data from the NIST beam neutron lifetime experiment described in reference \cite{Nic05} is available from the authors upon request.}, and reach a similar conclusion that the effect was $<$1 s in the neutron lifetime.
\par
Subsequent to the publication of Refs. \cite{Dew03,Nic05,Yue13} we conducted a detailed comparison of beam images using the dysprosium method and a Gadox (gadolinium oxysulfide) neutron camera \cite{GADOX,DISCLM} and found that the beam halo ($r > 8$ mm in Fig. \ref{F:BL1fig11}) in the neutron lifetime experiment was most likely an artifact. In retrospect our correction and uncertainty of \mbox{($-1.0 \pm 1.0$) s} was conservative.

\section{Proton Losses From the Detector Dead Layer}
\label{S:pDead}
The second issue raised by Serebrov, {\em et al.} concerns proton losses due to the silicon detector dead layer. This was an important systematic effect in the experiment. Depending on the detector used, 0.2 \% to 2 \% of incident protons backscattered from the dead layer and/or failed to deposit sufficient energy in the active volume to produce a countable pulse. We considered this effect carefully from the outset and designed the apparatus and experimental procedure to accommodate it.
\par
\begin{figure}
\centering
\includegraphics[width = 3.5in]{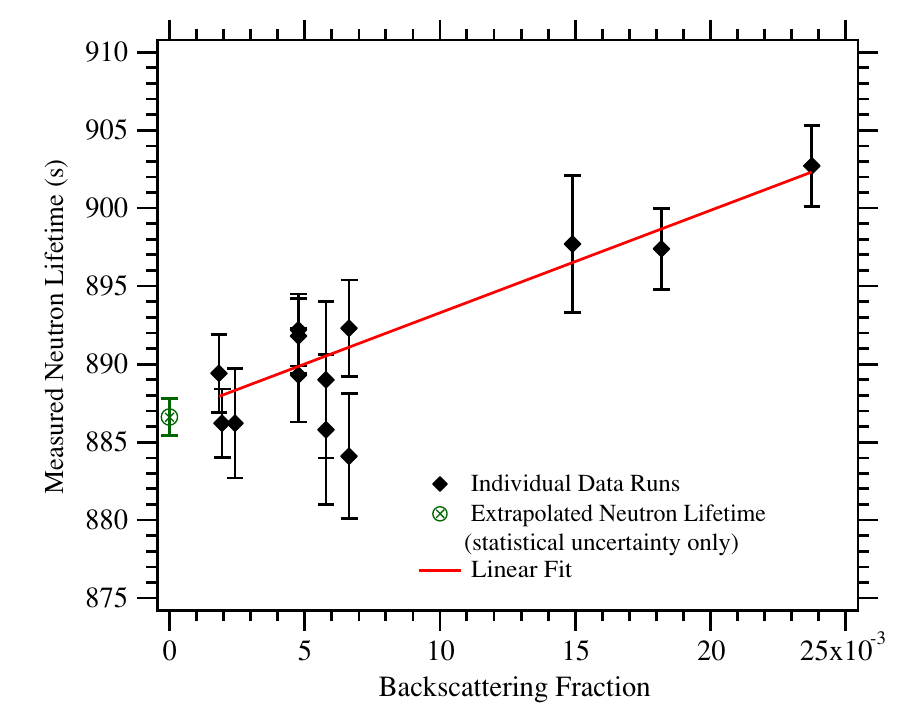}
\caption{\label{F:BL1fig20} A linear fit of the measured neutron lifetime versus the detector backscattering fraction. Reprinted from Ref. \cite{Nic05} (figure 11). The extrapolation
to zero backscattering gives the free neutron lifetime.}
\end{figure}

In the experiment, the neutron lifetime measurement was repeated using surface barrier detectors with different nominal thickness gold conducting layers, gold-free PIPS (Passivated Ion-implanted Planar Silicon) detectors, and different detector acceleration potentials (-32.5 kV to -27.5 kV). For each case we calculated the backscatter fraction both analytically and using the SRIM 2003 simulation package \cite{SRIM}. In our long experience modeling, measuring, and analyzing low energy proton spectra, we have found that SRIM predictions of the total backscatter probability are in good agreement with analytical modeling. However we have not succeeded in obtaining reliable predictions of the (energy, angle)-dependent backscatter spectrum from either SRIM or GEANT \cite{GEANT}. Therefore we felt we could not reliably correct the measured neutron lifetime values for backscatter effects using a detailed Monte Carlo simulation. Instead we followed a strategy of extrapolation. We plotted the measured neutron lifetime {\em vs.} calculated backscatter probability and extrapolated to zero backscatter as shown in Ref. \cite{Nic05}, figure 20, reproduced as figure \ref{F:BL1fig20} here. We expected, with good reason, the dependence of measured neutron lifetime on backscatter fraction to be monotonic, but we emphasize that we did not {\em a priori} assume, as suggested by Serebrov, {\em et al.}, a linear relationship. We extrapolated to zero using the simplest monotonic function that fit the data, which happened to be linear. We regard the true functional form of lifetime {\em vs.} backscatter fraction to be unknown. But given that we obtained a good fit to the data with a linear function, a more complicated function with additional parameters would not have improved the result. We agree with Serebrov, {\em et al.} that the energy spectrum of backscattered protons depends on the dead layer thickness and incident energy, and that this could in principle cause the measured lifetime {\em vs.} backscatter probability to be nonlinear. This was understood at the time of the 2005 experiment, but we did not observe evidence of such nonlinearity in our data at a statistically significant level.
\par
We stand by the results of this analysis, including our estimated uncertainty of 0.4 s due to this extrapolation, but we acknowledge that such a large extrapolation, over a range of measured lifetimes from 903 s to 886 s, is not ideal. It is a systematic weakness of the experiment; a much smaller extrapolation would be preferred. In the upcoming new beam neutron lifetime experiment BL3, the segmented silicon detector will be much larger and all backscattered protons that are reflected back by the electric field will strike the active region, unlike in our previous experiment where many missed, so the size of the required extrapolation will be much smaller.
\par
Serebrov, {\em et al.} produce a SRIM-based detailed Monte Carlo simulation of backscatter corrections in the NIST experiment. As explained above we do not consider such a course to be reliable. The authors of Ref. \cite{Ser21} lacked important details such as the experimental geometry and magnetic field shape. They apparently used only the nominal detector gold layer thicknesses. There is an additional layer of dead silicon that should be included, deduced by us using SRIM and experimental measurements of energy loss using protons and alphas. Table II in Ref. \cite{Ser21} implies a zero dead layer was used for the PIPS detectors, while in reality there is a small but significant silicon dead layer. Also they seem to have omitted the preacceleration of protons produced by the ramp potential in the trap. They conclude that our corrected neutron lifetime should be 2 s higher than our extrapolated result. Given the omissions and uncertainties of their method, we do not regard that conclusion to be valid.

\section{Proton Losses Due to Residual Gas Interactions}
\label{S:pRG}
Finally Serebrov, {\em et al.} consider the possible interactions of trapped protons with residual gas in the trap. First they make a simplified model of the vacuum environment of the trap as a vessel with cold walls located inside another vessel with warm walls (the outer vacuum system). They assume that residual gas flows from the outer vessel into the inner vessel, remaining in gas phase at thermal equilibrium with the walls in the two vessels. Therefore the molecular density in the inner vessel reaches equilibrium at $n = P/k\sqrt{T_1 T_2}$, where $P$ is the vacuum pressure in the outer chamber, $k$ is the Boltzmann constant, and $T_1$, $T_2$ are the vessel temperatures. Using $P = 10^{-9}$ mbar as the ion gauge pressure (actually the upper limit as the gauge was under range), $T_1 = 300$ K, and $T_2 = 4$ K, they obtain $n = 2.1\times 10^8$ cm$^{-3}$ inside the trap. Unfortunately this model omits the important effect of cryocondensation on the cold bore of the magnet, a crucial feature of the trap vacuum.
\par
\begin{figure*}
\centering
\includegraphics[width = 5in]{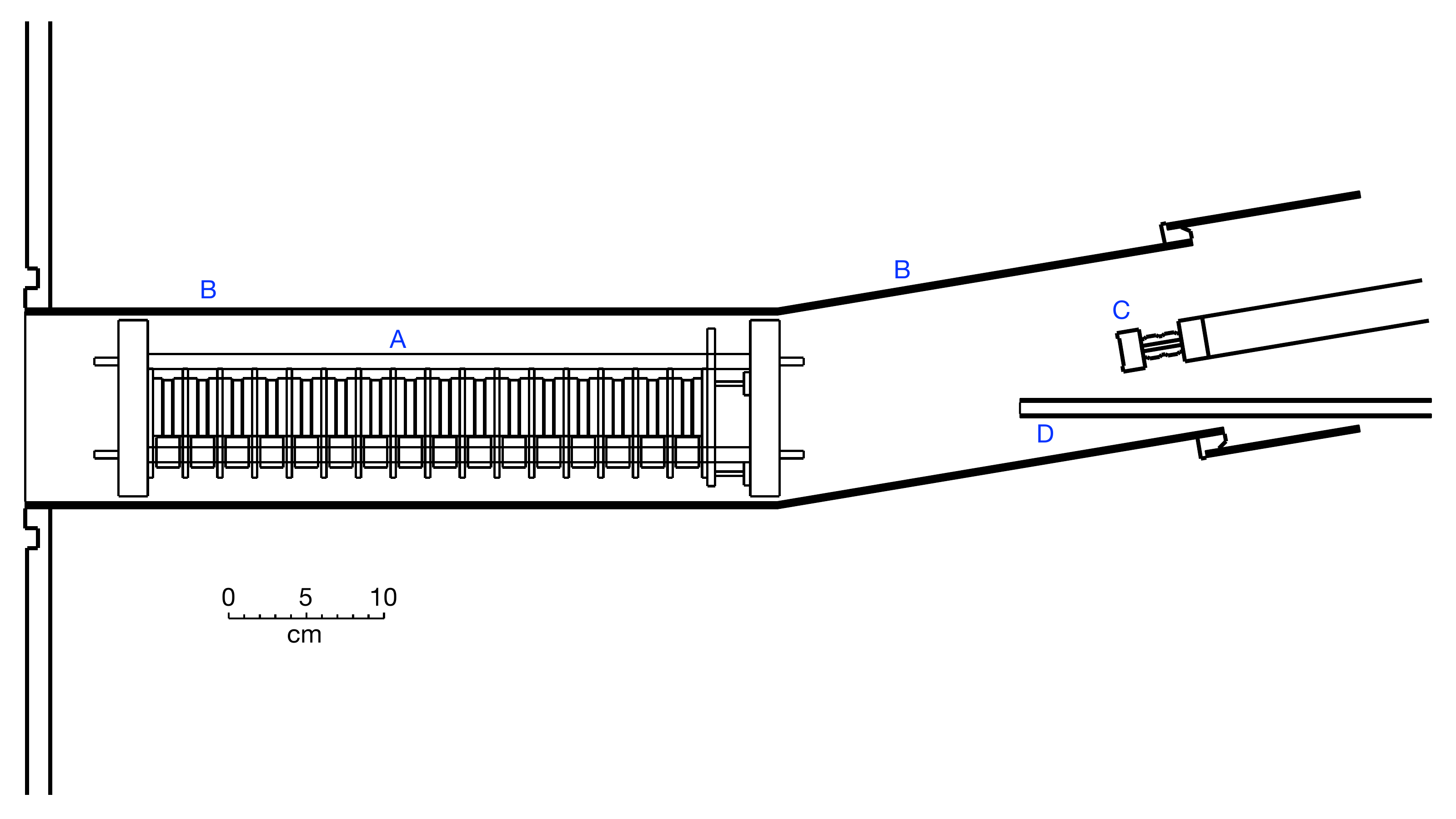}
\caption{\label{F:BL1arr} Arrangement of the proton trap apparatus in the NIST beam neutron lifetime experiment \cite{Dew03,Nic05}. A) proton trap; B) 8 K magnet bore; C) silicon proton detector; D) quartz neutron guide.}
\end{figure*}
\begin{figure*}
\centering
\includegraphics[width = 5in]{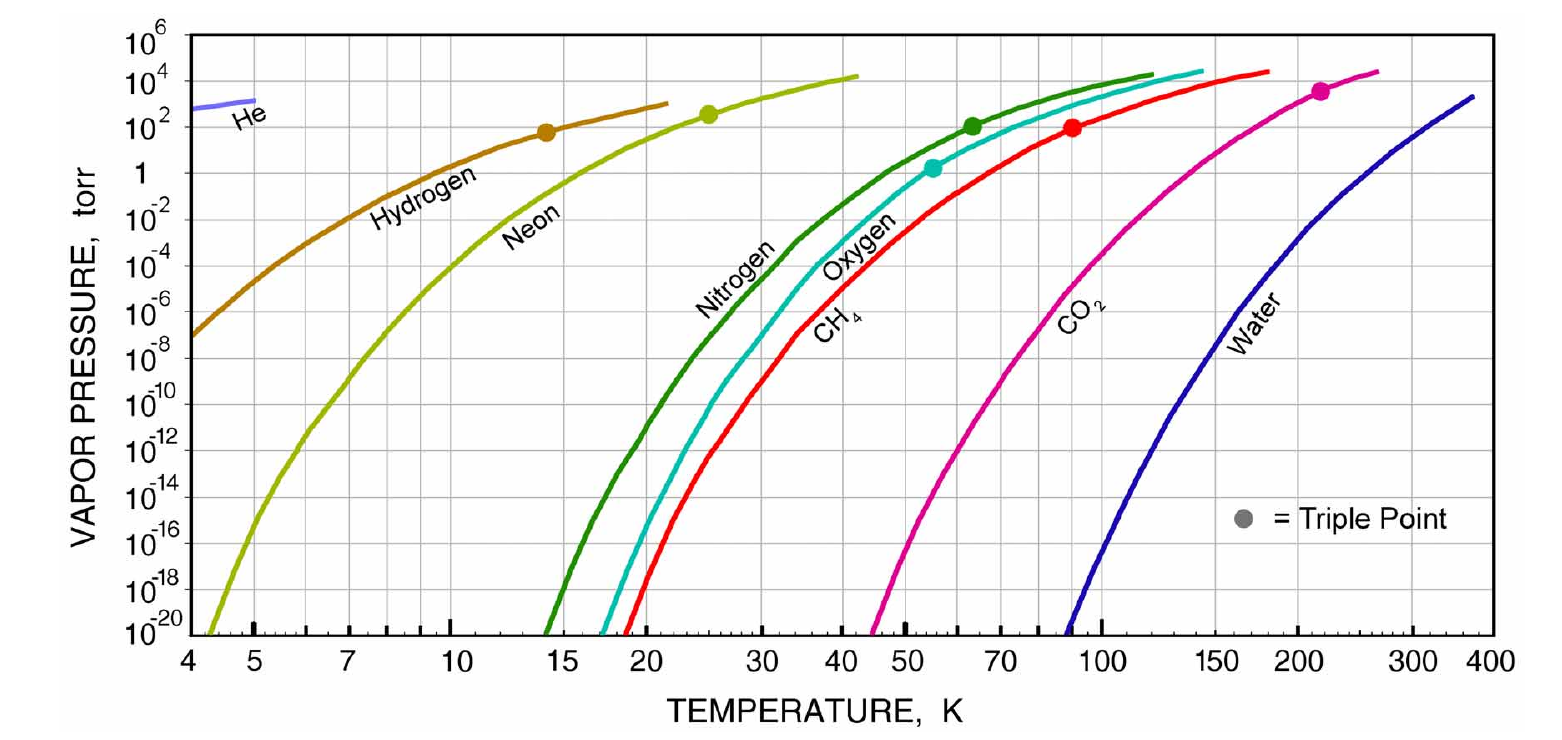}
\caption{\label{F:vaporP} Saturated vapor pressure of common gases as a function of temperature, from Ref. \cite{Ros19}.}
\end{figure*}

The arrangement of the bore, trap, and detector is shown in figure \ref{F:BL1arr}. The magnet bore was a 45 cm long, 12 cm inner diameter stainless steel tube in direct contact with the liquid helium bath. Its operational temperature was measured to be 8 K.  At this temperature the condensation coefficients of most gases are close to unity so residual gas will condense on the wall after just a few collisions, rather than remain in the gas phase and reach thermal equilibrium. The bore is effectively a cryopump. According to the theory of cryocondensation (see for example Refs. \cite{Bae87,Ohan}) the partial pressure of each gas component in the bore will reach equilibrium close to its saturation vapor pressure. Figure \ref{F:vaporP} shows a plot of saturation vapor pressure {\em vs.} temperature for a number of common gases. Other than hydrogen, helium, and neon the partial pressure and density of all residual gas components are predicted to be far lower than the estimate in Ref. \cite{Ser21}. There is no reason to expect neon in the vacuum system. Hydrogen is certainly present and in fact is the dominant residual gas. Helium is also a possibility due to its emission into the guide hall atmosphere from various cryogenic systems. Lacking important information about our vacuum system, Serebrov, {\em et al.} embark on a highly speculative discourse on the residual gas spectrum in our trap. They include the possibility of cryocondensation on the trap surfaces, which they assume to be in the range 20 K to 30 K. The trap was actually somewhat warmer, about 40 K, due to its weak conductive contact with the bore. At that temperature water will be pumped effectively but not other important gases such as air and methane. However they neglect to consider the far more powerful effect of cryopumping by the bore that surrounds the trap. 
\par
Residual gas interactions with trapped protons has been an important consideration in this experiment from the outset decades ago. We have extensively studied the potential effects of trapped proton interactions with hydrogen and helium, and there is recent theoretical work by others \cite{Byr19, Byr22}. The main concerns are charge exchange between a trapped proton and a H$_2$ molecule or He atom, leaving a H$_2^+$ or He$^+$ ion inside the trap, or binding of a trapped proton with a monatomic H atom, leaving a H$_2^+$. A key point is that both such trapped ions will be detected by the proton detector at a slightly later time relative to trapped protons. Due to the detector dead layer, the H$_2^+$ will appear at a lower, but detectable energy and the He$^+$ at slightly higher energy relative to the protons. We did not observe either of these in the 2005 experiment. In more recent data taken with the same apparatus, but with a different vacuum configuration, we believe we are able to observe trapped H$_2^+$ in certain vacuum conditions, but not at a level where they would significantly affect the neutron lifetime result given the way our data were analyzed. This was recently reported \cite{Cay22} and will be described more fully in an upcoming publication. 
\par
The trap timing plot shown in Ref. \cite{Ser21}, figure 16 is based on incorrect assumptions about our trap and detector geometry. Serebrov, {\em et al.} claim that the H$_2^+$ (shown in blue) would not be observed but would at the same time cause an overcounting of the background in region III. In reality, if such events were present at a statistically significant level, they would be visible and appear in region II in a correct analysis of the timing. The conclusion of Serebrov, {\em et al.} that this should be a $>$ 3 s correction, based on an analysis that lacked important details of the experiment, is incorrect.
\par
It is important to note that, hypothetically at least, circumstances could exist in the apparatus such that the limits of figure \ref{F:vaporP} at 8 K are exceeded. For example if residual gas molecules were ionized, they would be trapped by the magnetic field and may not interact with the cold surface of the bore. Gases originating inside the trap by outgassing or a virtual leak could allow much higher partial pressures within the trap compared to the outer bore region. A small fraction of gas molecules from the warm vacuum region will travel in ballistic trajectories that pass through the trap while missing the bore surface. If heavier molecules such as N$_2$ and CH$_4$ were present as trapped ions, they would lose most of their energy in the detector dead layer and be difficult to detect. Such possible residual gas effects continue to be an active area of investigation for the NIST beam experiment.
\par
Serebrov, {\em et al.} additionally consider elastic scattering interactions between trapped protons and residual gas molecules. We agree with their conclusion that this produces a negligible effect on proton counting efficiency at gas pressures that could have existed in the proton trap.

\section{Conclusions}
In Ref. \cite{Ser21} Serebrov, {\em et al.} discuss three potential systematic effects that they believe were not properly considered in the analysis of the NIST beam neutron lifetime experiment as described in Refs. \cite{Dew03,Nic05,Yue13}: 1) protons missing the active area of the proton detector; 2) losses due to the detector dead layer; and 3) residual gas effects. In particular they conclude that interactions between residual gas hydrogen and trapped protons could have led to a $>3$ s error in the result. 
\par
We have shown here that these effects were adequately analyzed with appropriate corrections and systematic uncertainties given in table V \cite{Nic05}. Regarding residual gas hydrogen, the resulting H$_2^+$ ions are detectable and produce a characteristic signature in our data. We saw no significant evidence of this effect in the 2005 experiment but have observed it in more recent data taken with the same apparatus but a different vacuum configuration. As recently reported \cite{Cay22}, even at the highest observed levels this systematic error in the neutron lifetime result would be less than 1 s. 
\par
A large class of potential systematic effects in the beam lifetime experiment, including residual gas interactions, will cause a loss of protons from the trap with a time scale of ms. Such effects would be made apparent by repeating the neutron lifetime measurement using a range of trapping times from 1 ms to 100 ms. This was not achievable in the original NIST experiment \cite{Dew03,Nic05,Yue13} due to trap instability at times over 10 ms. With improved trap stability such a program of measurements is an important goal for the current BL2 effort as well as the upcoming BL3 experiment.
\par
The neutron lifetime discrepancy is an important problem and we appreciate the effort made by Serebrov, {\em et al.} to examine our previous result and consider possible systematic effects. However, had they raised these points with us prior to their publication, we could have provided additional relevant information and clarifications. 

\section{Acknowledgements}
This work was supported by the National Institute of Standards and Technology (NIST), U.S. Department of Commerce; National Science Foundation grant PHY-2012395; and U.S. Department of Energy, Office of Nuclear Physics Interagency Agreement 89243019SSC000025 and grant DE-FG02-03ER41258. We acknowledge support from the NIST Center for Neutron Research, US Department of Commerce, in providing the neutron facilities used in this work.


\begin{thebibliography}{99}
\bibitem{Ser21} A.~P. Serebrov, M.~E.~Chaikovskii, G.~N.~Klyushnikov, O.~M.~Zherebtsov, and A.~V.~Chechkin, Phys. Rev. D {\bf 103}, 074010 (2021).
\bibitem{Dew03} M.~S.~Dewey, {\em et al.}, Phys. Rev. Lett. {\bf 91}, 152302 (2003).
\bibitem{Nic05} J.~S.~Nico, {\em et al.}, Phys. Rev. C {\bf 71}, 055502 (2005).
\bibitem{Yue13} A.~T.~Yue, {\em et al.}, Phys. Rev. Lett. {\bf 111}, 222501 (2013).
\bibitem{Rob51} J.~M.~Robson, Phys. Rev. {\bf 83}, 349 (1951).
\bibitem{Wie11} F.~E.~Wietfeldt and G.~L.~Greene, Rev. Mod. Phys. {\bf 83}, 1173 (2011).
\bibitem{Osh09} C.~M.~O'Shaughnessy, {\em et al.}, Nucl. Instr. Meth. {\bf A611}, 1 (2009). 
\bibitem{Ezh18} V.~F.~Ezhov, A.~Z.~Andreev, G.~Ban, B.~A.~Bazarov, P.~Geltenbort, A.~G.~Glushkov, V.~A.~Knyazkov, N.~A.~Kovrizhnykh, G.~B.~Krygin, O.~Naviliat-Cuncic, and V.~L.~Ryabov, JETP Lett. {\bf 107}, 671 (2018).
\bibitem{Pat18} R.~W.~Pattie, {\em et al.} (UCN$\tau$ collaboration), Science {\bf 360}, 627 (2018).
\bibitem{Gon21} F.~M.~Gonzalez, {\em et al.} (UCN$\tau$ collaboration), Phys. Rev. Lett. {\bf 127}, 162501 (2021).
\bibitem{Wil20} J.~T,~Wilson, {\em et al.}, Phys. Rev. Res. {\bf 2}, 023316 (2020).
\bibitem{Wil21} J.~T,~Wilson, {\em et al.}, arXiv:2011.07061 (2021).
\bibitem{GADOX} Ander Neo sCMOS camera viewing a Lexel Imaging 20 $\mu$m thick Gadoxysulfide scintillator.
\bibitem{DISCLM} Certain trade names and company products are 
mentioned in the text or identified in illustrations in order to 
adequately specify the experimental procedure and equipment used. In 
no case does such identification imply recommendation or endorsement 
by the National Institute of Standards and Technology, nor does it imply 
that the products are necessarily the best available for the purpose.
\bibitem{SRIM} J.~F.~Ziegler, computer code, “Stopping and Range of Ions in Matter" SRIM 2003, www.srim.org.
\bibitem{GEANT} GEANT 4, https://geant4.web.cern.ch.
\bibitem{Bae87} W.~G.~Baechler, {\em et al.}, Vacuum {\bf 37}, 21 (1987).
\bibitem{Ohan} J.~F.~O'Hanlon, {\em A User's Guide to Vacuum Technology}, 3/e, pp. 263--285, John Wiley \& Sons, USA (2003), ISBN 0-471-27052-0.
\bibitem{Ros19} R.~G.~Ross, in {\em Low Temperature Materials and Mechanisms}, ed. by Y.~Bar-Cohen, pp. 109-181, CRC Press, USA (2019), ISBN 0-367-87134-3.
\bibitem{Byr19} J.~Byrne and D.~L.~Worcester, J.~Phys. G: Nucl. Part. Phys. {\bf 46}, 085001 (2019).
\bibitem{Byr22} J.~Byrne and D.~L.~Worcester, Eur. Phys. J. A {\bf 58}, 151 (2022).
\bibitem{Cay22} J.~Caylor, Bulletin of the American Physical Society, APS April 2022 meeting, T12.01 (2022).

\end{thebibliography}
\end{document}